\documentstyle[epsf]{prhep97}

\begin{document}



\title{608: Jupiter Effect }
\author{Peter Filip (filip@savba.sk)}
\institute{Comenius University Bratislava, SK-84215, Slovakia}

\maketitle

\begin{abstract}
Specific results of the computer simulation of dilepton production from
expanding pion gas created in Pb+Pb 160 GeV/n collisions are
presented. Azimuthal asymmetry of dilepton pairs in non-central collisions
and interesting shape of the rapidity distribution of dilepton
pairs are predicted. These results are understood on theoretical level as a 
consequence of momentum and space asymmetries in the initial state of 
pion gas without any assumption of thermalization. Implication on
the production of dileptons in pre-hadronic phase of HIC is drawn.
\end{abstract}
\section{Introduction}

Goal of heavy ion collision (HIC) experiments is to reveal properties
of the compressed hot nuclear matter created during the collision of
heavy nuclei. Final momentum distributions of most abundant
particles - hadrons are however influenced
during the dilute and late freeze-out stage of heavy ion collision.
Interesting information about the early stage of the collision is in this case
hidden by subsequent collective effects of strongly interacting
hadrons.

Fortunately this is not valid for leptons or
photons. Distributions of these types of particles
can provide us with more direct information about the early
stages of the collision process.

In most of theoretical estimates for production of dileptons from
the hadronic matter created in HIC experiments an assumption about 
the equilibrium - thermalized stage of the collision is used.

Our study of dilepton production from the expanding pion gas is
not based on the assumption of thermalization.
Phenomena described
in next sections are generated also in the case of 1 collision per particle
approximation what means that the mean free path of particles participating
in mutual interactions is comparable with the size of the system.

In subsequent sections we report about two phenomena revealed by the computer
simulation of the dilepton production from expanding pion gas created in
Pb+Pb 160 GeV/n collisions: a) Azimuthal asymmetry of dilepton pairs in
non-central collisions, b) Rapidity distribution of dilepton pairs.

\section{Azimuthal Asymmetry of Dilepton Pairs}\label{sec4}
Azimuthal asymmetries of secondary produced particles have been
clearly identified in relativistic non-central collisions \cite{E877}.
This phenomenon is well understood as a consequence of collective
behavior of nuclear matter \cite{Ollitrault} or explained by the 
absorption of secondary produced particles in spectator parts of nuclei
\cite{HGR}. 
Recently also azimuthal asymmetries in
transverse momentum distributions of less abundant hadrons - $K$ mesons and
$\Lambda $ baryons \cite{HER} have been studied in HIC experiments.
However azimuthal asymmetries in transverse momentum distributions of
dileptons have not been addressed experimentally or theoretically
so far.

From theoretical point of view mechanisms
generating azimuthal asymmetries in transverse momentum distributions of
hadrons are not applicable for dileptons. Dileptons leave freely
collision volume after being produced without final state interactions or
absorption processes.

\vskip-0.4cm

\begin{figure}[h]
\centerline{\epsfxsize=9cm\epsffile{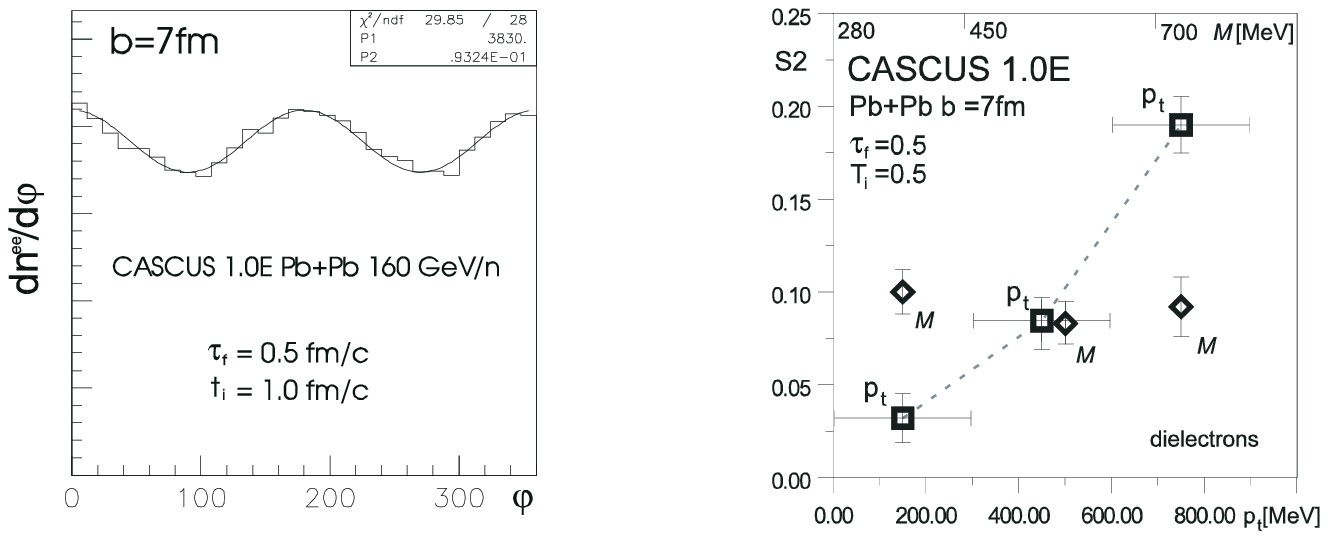}}
\vskip-0.3cm
\caption{Azimuthal asymmetry of dilepton pairs and its 
transverse momentum and invariant mass dependence for Pb+Pb 160 GeV/n $b=7$fm
events.}
\end{figure}

\vskip-0.4cm

In spite of this computer simulation \cite{nucl-th}
predicts significant second-order
asymmetry in transverse momentum distribution of dilepton pairs.
Result shown in Fig.1 is obtained by the simulation of Pb+Pb 160 GeV/n 
non-central $b$=7fm events. 
The fit of the histogram shown in Fig.1 to the function:
\begin{equation}
R(\phi )=S_0[1+S_2\cdot \cos (2\phi )]
\label{S2}
\end{equation}
gives numerical value of the asymmetry coefficient $S_2=0.093\pm 0.004$.
Corresponding value of $R_p$ parameter 
$R_p\approx \langle p_y^2 \rangle /\langle p_x^2 \rangle  = 1.202$.
Asymmetry of dilepton pairs is oriented {\it in} the reaction plane, it
increases with $p_t$ and does not depend on invariant mass
region.
Theoretical understanding of the origin of this asymmetry is sketched in 
Section 4 and studied more carefully in \cite{PhD}.

\section{Rapidity Distribution of Dilepton Pairs}\label{sec3}
Rapidity distribution of dilepton pairs produced via $\pi^+\pi^-$ annihilation
channel is determined by the rapidity distribution of momentum sum
$p_{\pi^+}+p_{\pi^-}$ of pions annihilating.
Result of simulation \cite{nucl-th} is shown in Fig.2 where 
also rapidity distribution
of pions participating in the rescattering process is shown.

For parameters of the simulation $\tau _f=0.5$ fm, $T_i=1.0$ fm 
number of collisions
per pion is close to 2.5 and minimum in the rapidity distribution of dileptons
is strong. For higher collision rates
(smaller values of $\tau _f $ and $T_i$) the minimum becomes weaker
and for 
$n_c=10$ coll./$\pi$ the minimum disappears.

\vskip-0.4cm
\begin{figure}[h]
\centerline{\epsfxsize=11cm\epsffile{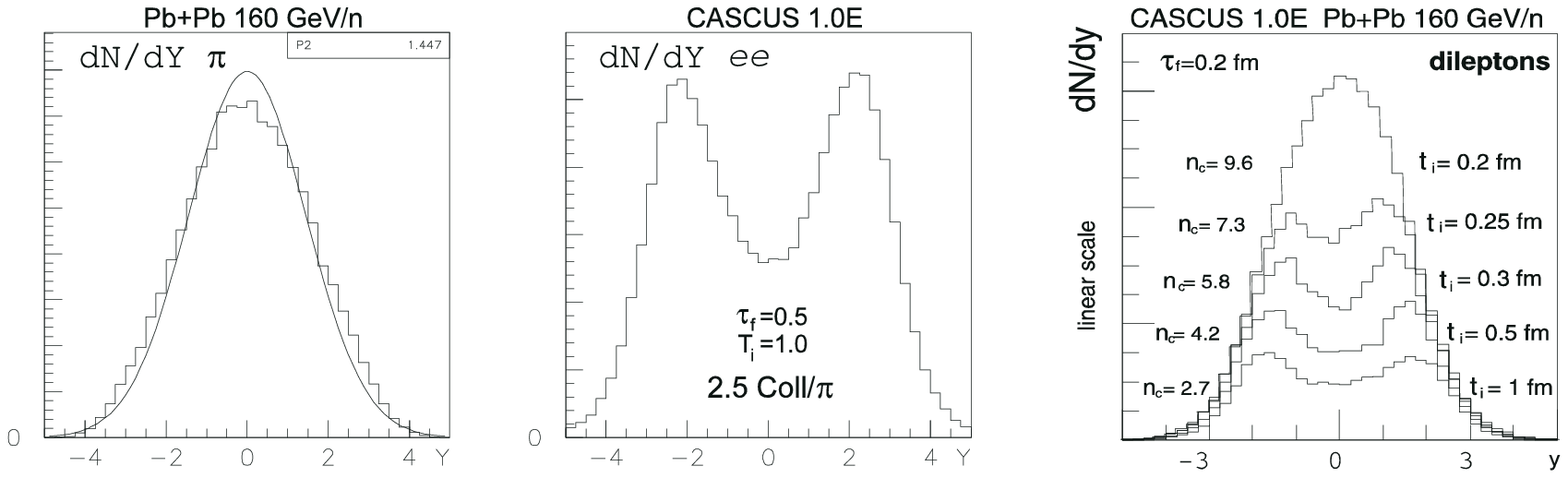}}
\vskip-0.3cm
\caption{Rapidity distribution of pions and dilepton pairs predicted
by the simulation of Pb+Pb 160 GeV/n collisions for different values of 
$\tau _f$, $T_i$.} 
\end{figure}
\vskip-0.4cm

From experimental point of view data on rapidity distribution
of dileptons produced in $p-A$ or $A-A$ collisions are rather
rare. It seems that the only published rapidity distribution
of dileptons which seem to originate from $\pi^+\pi^-$ annihilation
was obtained in pioneering experiments of
DLS collaboration at Bevalac accelerator in Berkeley \cite{DLS}.

\section{Theoretical understanding of results}
Azimuthal asymmetry of dilepton pairs and rapidity
distribution of dileptons obtained in the simulation can be understood
without the assumption of thermalization.
Initial transverse momentum distribution of pions 
is azimuthally symmetrical - constant 
in the simulation. Therefore azimuthal asymmetry of dileptons is not
generated by transverse momentum asymmetry of pions.
It is a consequence of the initial spatial distribution
of pions in transverse plane which is asymmetrical
in the case of non-central collisions.

\vskip-0.5cm
\begin{figure}[h]
\centerline{\epsfxsize=11cm\epsffile{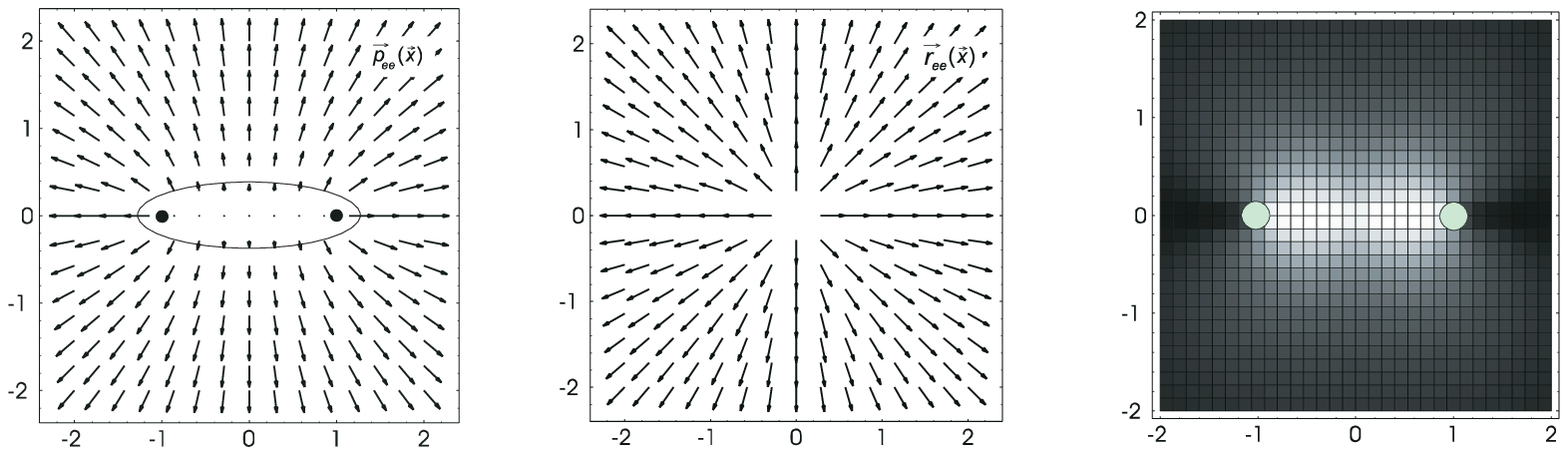}}
\vskip-0.3cm
\caption{
Vector field $p_{ee}(x,y)$ of dilepton pairs created by $\pi^+\pi^-$ 
annihilation of
pions emitted from two sources. The two-point source serves as a
rough approximation of the asymmetrical shape of the initial distribution of 
pions in transverse plane.  
Density-plot of the difference
$|p_{ee}(x,y)-r_{ee}(x,y)|$ ($r_{ee}(x,y)$ is radial field) 
shows that the asymmetry 
is generated between emission points - in the volume of source.
}
\end{figure}
\vskip-0.5cm

Rapidity distribution of dilepton pairs can be explained as a consequence
of the asymmetry of pions in momentum space. Average transverse momentum
of pions in Pb+Pb 160 GeV/n collisions is much smaller compared to average 
longitudinal momentum of pions.
This strongly influences distribution of pion-pion collisions in rapidity.
In Fig.4
we show distribution of annihilation points of $\pi ^+\pi ^-$ pairs emitted
from two-point source oriented parallel to the
beam direction. Momentum distribution     
of emitted pions is asymmetrical: $<|p_l|>\approx 2 <p_t>; <p_t>=380$ MeV.
Annihilation events in the middle of the sources
are suppressed due to the shape of $\pi^+\pi^- \rightarrow e^+e^-$ 
cross section which is peaked at $M=770$MeV (for detailed description 
see \cite{PhD}).

\vskip-0.5cm
\begin{figure}
\centerline{\epsfxsize=8cm\epsffile{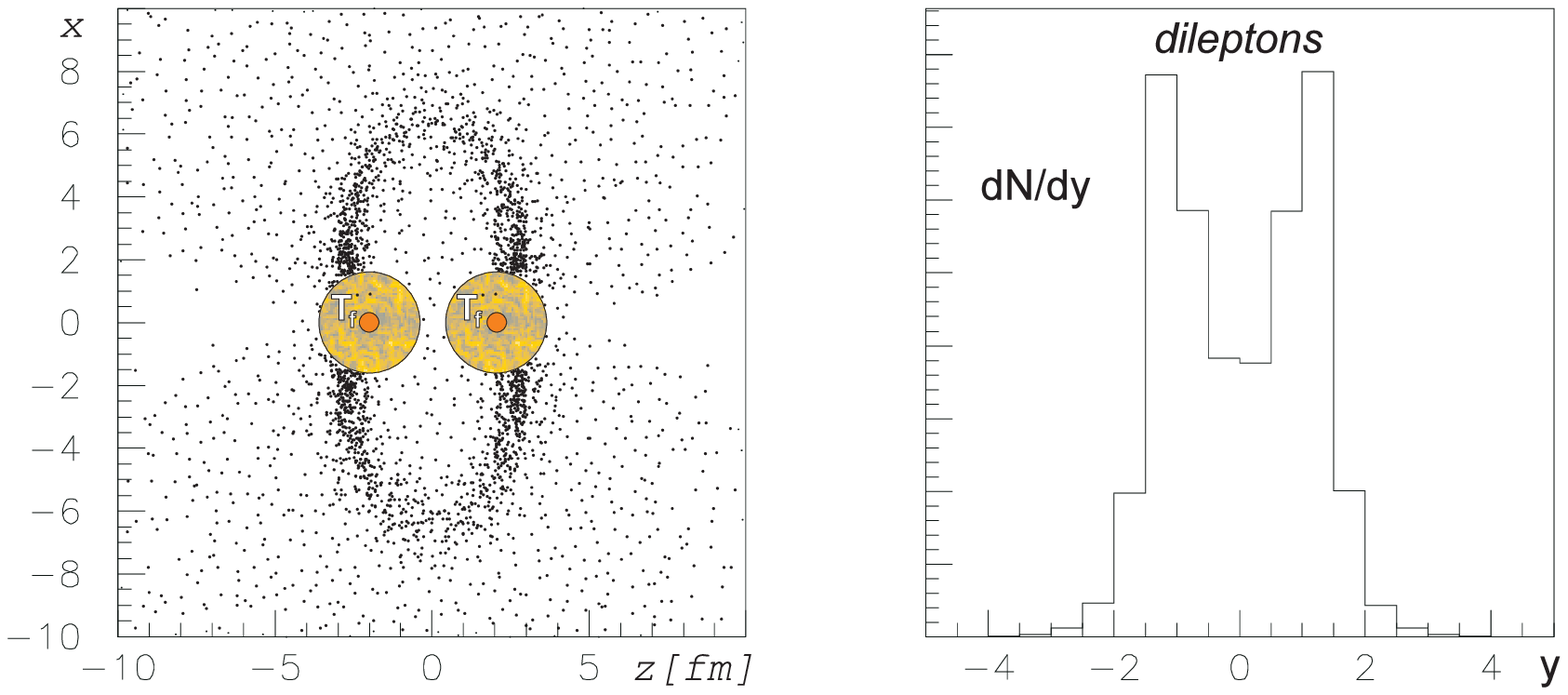}}
\vskip-0.3cm
\caption{Distribution of annihilation points of pions emitted from two-point
source described in the text and rapidity distribution of the generated
dilepton pairs.
Formation time \cite{Pisut}
excludes annihilations close to emission points.
}
\end{figure}
\vskip-1.5cm

\  

\section{Conclusions}

\vskip-0.3cm

Phenomena described in preceding sections are generated in  
pre-equilibrium stage of the pion gas expansion. 
Similar effects might occur also in parton gas possibly
created in heavy ion collision experiments. In this case the azimuthal
asymmetry of dilepton pairs in the invariant mass region 2-5 GeV might
be considered as a signature of secondary collisions \cite{Ruus} among
partons. 
Gas-like behavior
of the system of partons \cite{Jupiter} created in 
HIC experiments is worth
of further theoretical and experimental study.

\end{document}